\documentclass[final,5p,times,twocolumn,longtitle]{elsarticle}

\usepackage{graphicx}

\usepackage{amssymb}

\usepackage{xcolor}
\definecolor{darkgreen}{named}{green}
\definecolor{darkblue}{named}{blue}
\definecolor{darkred}{named}{red}
\definecolor{grey}{named}{gray}
\definecolor{middlegray}{rgb}{0.5,0.5,0.5}
\definecolor{lightgray}{rgb}{0.8,0.8,0.8}
\definecolor{orange}{rgb}{0.8,0.3,0.3}
\definecolor{yac}{rgb}{0.6,0.6,0.1}

\usepackage[]{listings}
\lstdefinestyle{xml_sty}{
language=XML,
commentstyle=\itshape\color{orange},
keywordstyle=\bfseries\color{orange},
tagstyle=\bfseries\color{darkblue},
stringstyle=\color{darkred},
extendedchars=true,
basicstyle=\scriptsize\ttfamily,
tabsize=2,
numbers=left,
numberstyle=\tiny,
breakautoindent = true,
breakindent = 2em,
breaklines = true,
showspaces=false, 
showtabs=false, 
showstringspaces=false,
frame=single,
tabsize=2,
captionpos=b
}

\journal{Nuclear Instruments and Methods in Physics Research, A}

\begin{document}

\begin{frontmatter}



\title{Advanced functionality for radio analysis in the Offline software framework of the Pierre Auger Observatory}

\author[72]{P.~Abreu}
\author[55]{M.~Aglietta}
\author[88]{E.J.~Ahn}
\author[17,88]{I.F.M.~Albuquerque}
\author[31]{D.~Allard}
\author[1]{I.~Allekotte}
\author[91]{J.~Allen}
\author[93]{P.~Allison}
\author[65]{J.~Alvarez Castillo}
\author[79]{J.~Alvarez-Mu\~{n}iz}
% ...
\address[1]{Centro At\'{o}mico Bariloche and Instituto Balseiro (CNEA-UNCuyo-CONICET), San Carlos de Bariloche, Argentina}
% ...
\address[17]{Universidade de S\~{a}o Paulo, Instituto de F\'{\i}sica, S\~{a}oPaulo, SP, Brazil}
% ...
\address[31]{Laboratoire AstroParticule et Cosmologie (APC),Universit\'{e} Paris 7, CNRS-IN2P3, Paris, France}
% ...
\address[55]{Istituto di Fisica dello Spazio Interplanetario (INAF),Universit\`{a} di Torino and Sezione INFN, Torino, Italy}
% ...
\address[65]{Universidad Nacional Autonoma de Mexico, Mexico, D.F.,Mexico}
% ...
\address[72]{LIP and Instituto Superior T\'{e}cnico, Lisboa, Portugal}
% ...
\address[79]{Universidad de Santiago de Compostela, Spain}
% ...
\address[88]{Fermilab, Batavia, IL, USA}
% ...
\address[91]{New York University, New York, NY, USA}
% ...
\address[93]{Ohio State University, Columbus, OH, USA}
% ...


\fntext[fn1]{Deceased}
\fntext[fn2]{at Konan University, Kobe, Japan}

\cortext[cor]{Corresponding author: auger\_pc@fnal.gov}

\begin{abstract}

The advent of the Auger Engineering Radio Array (AERA) necessitates the development of a powerful framework for the analysis of radio measurements of cosmic ray air showers. As AERA performs ``radio-hybrid'' measurements of air shower radio emission in coincidence with the surface particle detectors and fluorescence telescopes of the Pierre Auger Observatory, the radio analysis functionality had to be incorporated in the existing hybrid analysis solutions for fluoresence and surface detector data. This goal has been achieved in a natural way by extending the existing Auger Offline software framework with radio functionality. In this article, we lay out the design, highlights and features of the radio extension implemented in the Auger Offline framework. Its functionality has achieved a high degree of sophistication and offers advanced features such as vectorial reconstruction of the electric field, advanced signal processing algorithms, a transparent and efficient handling of FFTs, a very detailed simulation of detector effects, and the read-in of multiple data formats including data from various radio simulation codes. The source code of this radio functionality can be made available to interested parties on request.

\end{abstract}


\end{frontmatter}


\section{Introduction}

Forty years after the initial discovery of radio emission from extensive air showers \citep{JelleyFruinPorter1965}, the CODALEMA \citep{ArdouinBelletoileCharrier2005} and LOPES \citep{FalckeNature2005} experiments have re-ignited very active research activities in the field of radio detection of cosmic ray air showers. Nowadays, the field is in a phase of transition from first-generation experiments covering an area of less than 0.1 km$^2$ to large-scale arrays of tens of km$^2$. In particular, the Auger Engineering Radio Array (AERA) \citep{HuegePisa2009} will complement the southern site of the Pierre Auger Observatory \citep{AugerNIM2004} with $161$ autonomous radio detector stations covering an area of $\approx 20$~km$^2$.

One particular merit of the Pierre Auger Observatory is its hybrid mode of observation, which uses coincident detection of extensive air showers with both optical fluorescence telescopes (FD) and surface particle detectors (SD) to gain in-depth information on the measured air showers. Consequently, the analysis software has to support complete hybrid processing and interpretation of the data. This requirement is fulfilled by the Auger Offline software framework \citep{ArgiroOffline2007}. To take full advantage of the radio data taken in the hybrid environment of the Pierre Auger Observatory, it is clear that also radio analysis functionality, which has so far been existing in a separate software package \citep{FliescherArena2008}, had to be included in this hybrid analysis framework.

In this article, we describe how we have therefore built advanced radio analysis functionality into the Auger Offline software framework. The general structure of the radio implementation in the Offline framework will be discussed in section \ref{sec:structure}. A number of innovative features have been realized in this context for the very first time. These and other highlights will be discussed in section \ref{sec:highlights}. Finally, in section \ref{sec:analysis} we demonstrate how the advanced radio functionality embedded in the Offline framework can be used to carry out a complete detector simulation and event reconstruction on the basis of a simulated radio event.

\section{Embedding radio functionality in the Offline framework} \label{sec:structure}

The Offline framework has a clear structure to allow for easy maintenance and ongoing shared development over the whole life-time of the Pierre Auger Observatory \citep{ArgiroOffline2007}. In particular, there is a clear separation between the internal representation of the \emph{Detector} and the \emph{Event}. The \emph{Detector} provides access to all of the relevant detector information such as the positions of detector stations in the field, the hardware associated with these stations, etc. The \emph{Event} data structures in contrast hold all of the data applying to a specific event, such as ADC traces, but also reconstructed quantities such as the event geometry. There is no direct connection between these two entities. Instead, analysis \emph{Modules} use the defined interfaces of both the \emph{Detector} and \emph{Event} data structures to carry out their specific analysis tasks. No interface exists either between separate analysis modules, which can only propagate their results through the \emph{Event} data structure. This ensures that dependencies between analysis modules are kept to a minimum and facilitates the replacement of individual modules with alternative implementations, thereby providing a very high degree of flexibility.

\begin{figure*}[h!]
\centering
\includegraphics[width=0.8\textwidth]{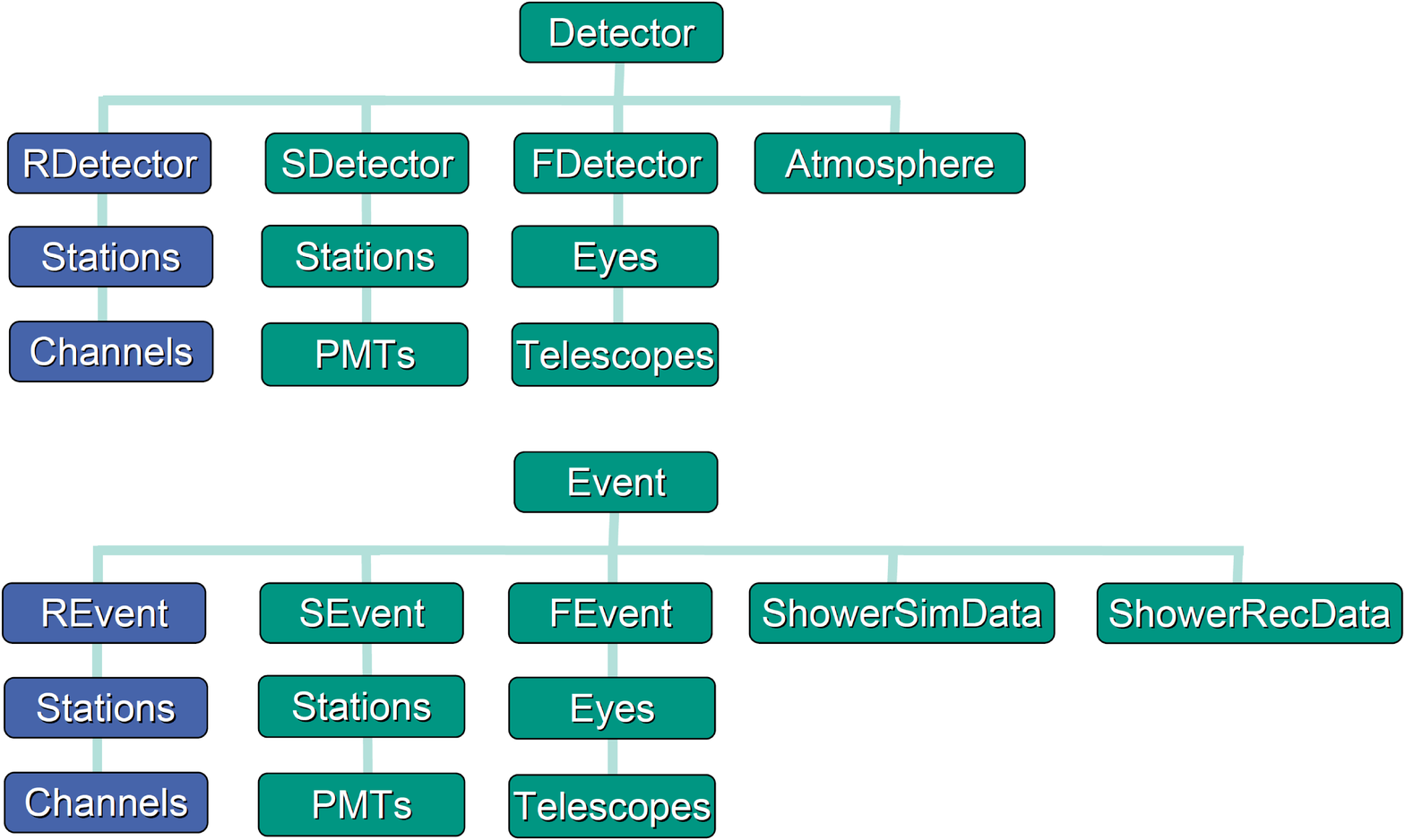} 
\caption{Within Offline, the detector and event data structures are clearly separated. Both data structures have been complemented with the analogous classes for radio detection (marked in blue). In both the \emph{RDetector} and the \emph{REvent}, classes for \emph{Stations} and \emph{Channels} are present. Those in the detector data structures provide access to the detector description, the ones in the event data structures store data applying to specific radio events.\label{fig:radioembedding}}
\end{figure*}

Clearly, the radio analysis functionality had to be implemented following the same philosophy. The hierarchical implementation of the radio parts of both the \emph{Detector} and \emph{Event} classes in addition to the existing FD- and SD-specific classes is depicted in Fig.\ \ref{fig:radioembedding}. In analogy to the hierarchy of \emph{Stations} and \emph{PMTs} in the SD functionality, the implementation of the radio data structures has been divided into \emph{Stations} and \emph{Channels}. A \emph{Station} represents one location in the field at which the electric field of the radio waves is measured. Data stored at \emph{Station} level therefore represents the physical electric field devoid of any detector influence except for the location (and limited observing bandwidth) of the \emph{Station}. In contrast, \emph{Channels} represent the individual antenna channels at which the ``raw'' measurement is performed by an ADC digitizing voltages. This clear separation between \emph{Channels} and \emph{Stations} is a very powerful concept and is original to the radio implementation in Offline. We will discuss its significance, among other highlights, in the following section.

\section{Highlights of the radio analysis functionality} \label{sec:highlights}

The radio functionality in the Offline framework provides a number of unique features facilitating an advanced radio data analysis. In this section, we will describe some of these highlights.


\subsection{Clear separation of Channel- and Station-levels} \label{sec:levelseparation}

When analyzing radio data, one is faced with two different ``levels''. The \emph{Channel} level is defined by the detector channels acquiring the raw data. These data consist of time-series of samples digitized with a sampling rate adequate for the frequency window of interest. Each sample denotes a scalar quantity such as an ADC count recorded by the channel ADC. Low-level detector effects such as the correction for the frequency-dependent response of cables, filters and amplifiers are treated on this level for each \emph{Channel} individually. Likewise, detector-related studies such as the evaluation of trigger efficiencies would be typically performed on \emph{Channel} level. When reading in measured data files, the raw data (ADC counts) are filled into the appropriate \emph{Channel} data structures.

In contrast, the \emph{Station} level is defined by the physical electric field present at a given location in the field, stored as a time-series of three-dimensional vectors. It is on \emph{Station} level that radio pulses are identified and quantified, before a geometry reconstruction of the given event is performed. Once the event reconstruction has been completed, the data at \emph{Station} level no longer have any dependence on the detector characteristics, except for the location and limited observing bandwidth of the measurement. A reconstruction of the electric field on the \emph{Station} level is therefore suited best for a comparison of radio measurements of different experiments, as well as for the comparison of radio measurements with corresponding simulations. Since simulated electric field traces provided by radio emission models also represent physical electric fields independent of a given detector, they are read in on the \emph{Station} level.

Analysis modules in Offline usually work on either \emph{Channel} or \emph{Station} level, and typically it is very clear which analysis step has to be performed on which level. The transition between the two levels is performed by applying the characteristics of the antennas associated to each of the \emph{Channels}. This transition can be employed in both directions, from \emph{Station} to \emph{Channel} or vice-versa. The transition from a \emph{Station} to the associated \emph{Channels} is typically performed to calculate the response of the individual detector \emph{Channels} to an electric field provided by simulations. The opposite transition is required when reconstructing the three-dimensional electric field vector from the data recorded by the (typically) two measurement channels in the field. This reconstruction will be further discussed in section \ref{sec:efieldreco}.

\subsection{Read-in from different data sources}

The \emph{Event} data structures are complemented with reader functionality to populate them with data available in one of several file formats for both experimentally measured data and simulated radio event data. Due to its wealth of supported formats and the possibility of easy extension with new formats, the radio functionality in Offline therefore provides very powerful functionality to compare data and simulations from different sources, which again is an original feature usually not found in the analysis software suites developed in the contexts of other experiments. At the time of writing, the following data formats are supported. For experimental data:

\begin{itemize}
\item{measurement data from two different prototype setups situated at the Balloon Launching Station of the Pierre Auger Observatory \citep{CoppensArena2008,RevenuArena2008}}
\item{measurement data from AERA \citep{HuegePisa2009}}
\end{itemize}
For simulation data, the following formats are currently readable:
\begin{itemize}
\item{simulation data from MGMR \citep{ScholtenWernerRusydi2008}}
\item{simulation data from REAS2 and REAS3 \citep{HuegeUlrichEngel2007a,LudwigHuege2010}}
\item{simulation data from ReAIRES \citep{DuVernoisIcrc2005}}
\end{itemize}


\subsection{Modular approach}

The strict interface design of the \emph{Detector}, the \emph{Event} and the analysis modules allows for a very modular implementation of radio analysis functionality. As the analysis modules are the part of the code typically the most exposed to the end-user, their interface has been kept relatively simple. End-users developing analysis functionality for Offline therefore only need relatively basic proficiency in C++.

An analysis application within Offline is defined through a ``module sequence'' in XML syntax, an example of which is listed in section \ref{sec:analysis}. In such a module sequence, analysis modules are chained in a meaningful sequence to perform a specific analysis task. The individual modules do not communicate directly with each other, but only share data through the \emph{Event} data structures. Consequently, modules can easily be removed, replaced or rearranged within a module sequence. This does not require recompilation of the source code. Additionally, each module can be configured individually through XML files.


\subsection{Transparent FFT handling}

Radio analyses typically apply algorithms both on time- and frequency-domain data. As a consequence, they heavily rely on fast Fourier transforms (FFTs). The Offline framework has thus been extended with FFT functionality based on the FFTW library \citep{FFTW}. A special feature of this implementation is that FFTs are handled completely transparently in the background. The user does not need to invoke FFTs manually.

\begin{figure*}[t!]
\centering
\includegraphics[width=0.8\textwidth]{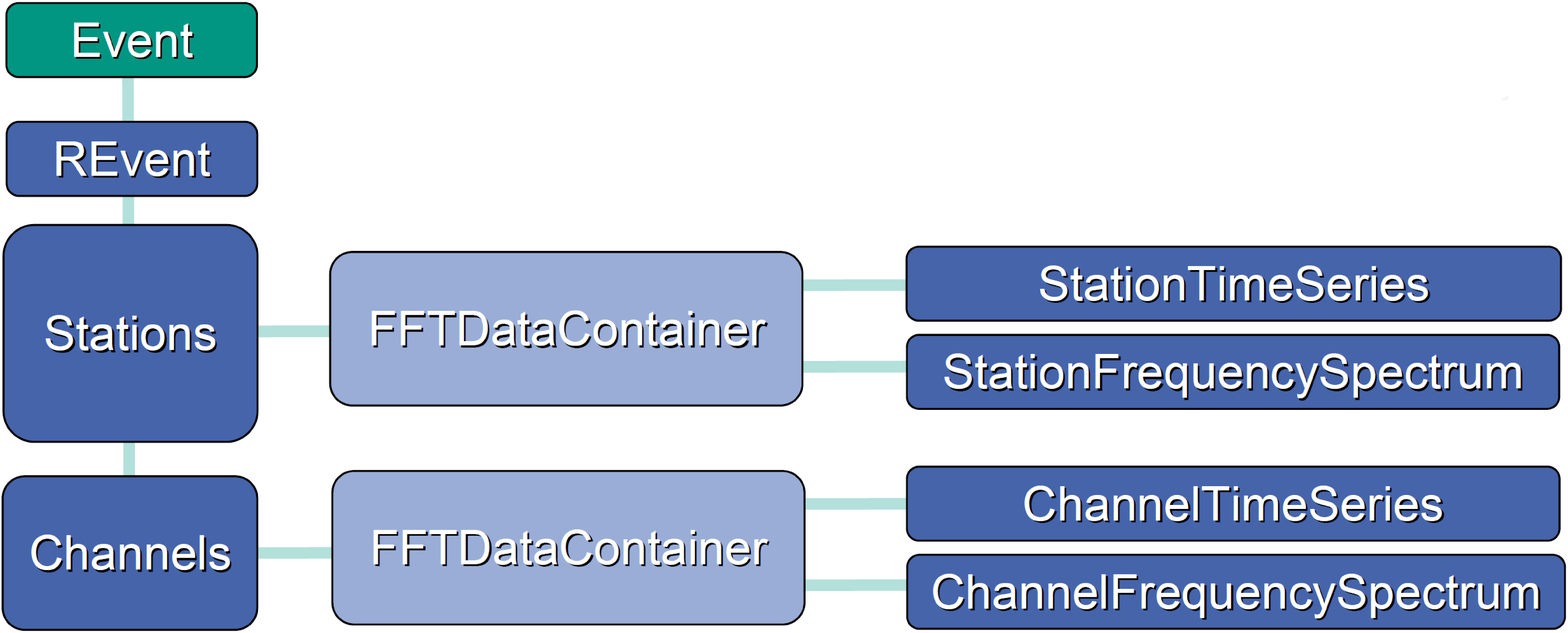}
\caption{At both the \emph{Station} and \emph{Channel} levels of the \emph{REvent}, data structures exist to store time-series and frequency-domain data. These are encapsulated in \emph{FFTDataContainers} which transparently and efficiently handle all necessary FFTs without explicit interaction from the end-user.\label{fig:offlinestructure:datacontainers}}
\end{figure*}

This is realized by the use of \emph{FFTDataContainers} as illustrated in Fig.\ \ref{fig:offlinestructure:datacontainers}. These containers encapsulate both the time- and frequency-domain representations of radio data on the \emph{Channel} and \emph{Station} levels. The user can access both the time-domain and frequency-domain data at any time. The \emph{FFTDataContainer} keeps track of which representation has been changed last and whether an FFT has to be performed or not before the data requested by the user are returned. All data are passed by reference and changed in place, so that even traces with an extreme length can be handled efficiently.

As a consequence of this design, the user can simply chain analysis modules working in any of the two domains without worrying which domain has last been worked on. (There is a performance benefit when grouping modules working in the same domain together, but it is not very significant.)


\subsection{Advanced analysis modules}

A number of analysis modules performing recurring steps in advanced radio analysis pipelines are available by default. They can easily be included or excluded from module sequences as needed:

\begin{itemize}
\item{modules applying bandpass filters to the \emph{Channel} and \emph{Station} levels}
\item{a module performing an up-sampling of under-sampled data}
\item{a module resampling data to a different time-base}
\item{a module suppressing narrow-band radio frequency interference through a ``median filter''}
\item{a module performing an enveloping of time traces via a Hilbert transform}
\item{a module determining timing differences between different antenna stations from the reference phases of a beacon transmitter}
\item{modules applying a windowing function (e.g., Hann window) to the \emph{Channel} and \emph{Station} levels}
\end{itemize}


\subsection{Detailed simulation of the detector response}

\begin{figure}[ht]
  \centering
  \includegraphics[width=0.45\textwidth]{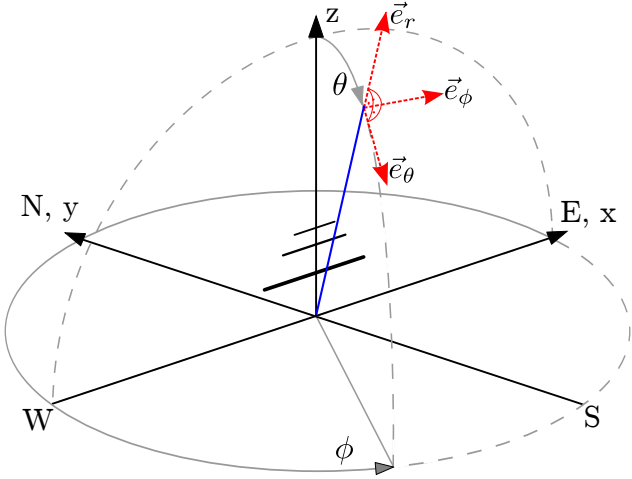}
  \caption{The antenna characteristics are defined in a spherical coordinate system with the antenna in its center. The effective antenna height $\vec{H}_{\mbox{eff}}$ for emission coming from a given arrival direction is decomposed into the components along the unit vectors $e_{\theta}$ and $e_{\phi}$. These local (i.e., arrival direction-dependent) unit vectors lie in the plane perpendicular to the Poynting vector, which aligns with the dashed blue line in this figure for the incoming direction defined by $\theta$ and $\phi$. As electromagnetic waves in air have no electric field component along the Poynting vector, this representation is complete. The antenna height itself is a frequency-dependent, complex quantity, the amplitude of which denotes the gain of the antenna, while the phase provides information about signal delays and dispersion.}
  \label{fig:AntennaCoordSys}
\end{figure}

When comparing measured data to simulated radio pulses from various models, it is required to perform a detailed simulation of the effects introduced by the various detector components. This encompasses in particular:
\begin{itemize}
\item{the complex response (impulse response defined by the frequency-dependent amplitudes and phases\footnote{A full transport matrix representing the transmission in forward direction, transmission in backward direction, as well as the reflections on the input and output could be implemented for a more detailed description. For the moment, however, we assume that impedance matching in the experimental setup is sufficiently good so that transmission in the forward direction describes the detector response with good precision.}) of all the analogue components (cables, filters, amplifiers) in each individual channel}
\item{the frequency- and direction-dependent complex gain (or ``effective antenna height'') of the antenna connected to each individual channel (cf.\ Fig.\ \ref{fig:AntennaCoordSys})}
\item{effects introduced by the sampling of the data with a given sampling rate}
\item{saturation effects occurring at the ADCs}
\item{effects introduced by the layout of the array, including geometric effects occurring on large scales due to the curvature of the Earth}
\end{itemize}
All of this functionality has been implemented in the Offline framework. At the moment, detector description data are provided as XML files. Later, a transition to MySQL or SQLite databases is foreseen and can be performed transparently. The complex response of individual \emph{Channels} is provided via a \emph{ResponseMap} detailing the hardware elements comprising each individual channel. The overall response of each channel is then calculated on-the-fly from the tabulated responses of each individual hardware component listed in the \emph{ResponseMap}. A caching mechanism ensures that overall responses are only recalculated when needed.


\subsection{Vectorial E-field reconstruction} \label{sec:efieldreco}

The physical electric field is a three-dimensional, vectorial quantity. When comparing results from different experiments or experimental results and radio emission models, the electric field is the quantity of choice, as in principle it has no dependence on the detector (except for the location at which it was measured and the limited observing bandwidth). Most radio detectors, however, are only equipped with two channels per position in the field, typically measuring the east-west and north-south linear polarization components. In other words, they only measure a projection of the three-dimensional electric field to the horizontal plane. In such a setup, two \emph{Channels} are available at each detector station, one connected with an east-west-aligned antenna and one connected with a north-south-aligned antenna. The (scalar) response of each individual channel to the incoming electric field can be calculated as the dot product between the electric field vector and the vectorial representation of the effective antenna height. The effective antenna height is related to the gain of the antenna and depends on the arrival direction and frequency (cf.\ Fig.\ \ref{fig:AntennaCoordSys}). Consequently, this calculation is best done in the frequency domain.

The more difficult problem is the inverse calculation: the reconstruction of the three-dimensional electric field vector from the two-dimensional measurement. This inversion is possible if the arrival direction of the electromagnetic wave is known, because electromagnetic waves in the atmosphere constitute transverse waves, the electric field of which lies in a plane perpendicular to the direction of propagation.\footnote{At the moment, we assume that the propagation direction of radio emission from extensive air showers can be approximated well with the shower axis.}

The antenna characteristics needed to reconstruct the vectorial electric field on the \emph{Station} level depend on the arrival direction. This arrival direction, however, is not available until after the reconstruction on the basis of the \emph{Station} level three-dimensional electric field. Therefore, an iterative approach starting with a reasonable initial arrival direction is performed in the radio analysis in Offline. The reconstructed arrival direction quickly converges to its final value, and the vertical component of the electric field can be reconstructed from the two-dimensional measurement (cf.\ section \ref{sec:analysis}). This reconstruction scheme is truly original to the Offline radio functionality.


\subsection{Advanced data output}

\begin{figure}
\begin{center}
\includegraphics[width=0.5\textwidth]{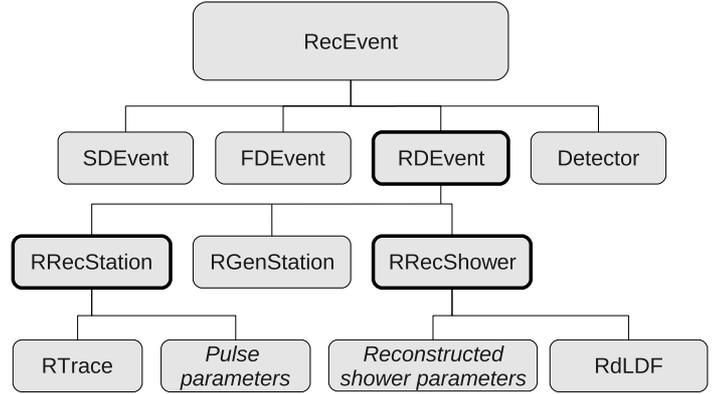} 
\end{center}
\caption{Hierarchical structure of the radio-related quantities stored in ADST files. The individual classes constitute logical entities present in the \emph{Event} data structures, storing for example information on Monte Carlo generated quantities in \emph{RGenStation} and the reconstructed radio lateral distribution function in \emph{RdLDF}.\label{RecEvent-class}}
\end{figure}

End-users performing a higher-level analysis usually do not need access to all the raw data of each individual event. They rather need information about reconstructed quantities derived by the low-level analysis pipeline. A data format to store such quantities reconstructed by an Offline-based analysis is the Advanced Data Summary Tree (ADST). ADST files hold relevant data for FD, SD and now also radio-reconstructed quantities. The structure of the radio data within the ADST files is depicted in Fig.\ \ref{RecEvent-class}.
In addition to accessing the content of ADST files directly from end-user analysis programs, a graphical user interface exists for browsing the contents of ADST files and visualizing the included events. This \emph{EventBrowser} has also been complemented with radio-specific functionality, so that also the radio part of the event (such as traces, spectra, a map of the array and event geometry, ...) can be visualized in a user-friendly and intuitive fashion.


\subsection{Easy extension to hybrid analysis}

One of the main motivations for including radio functionality in the Offline framework was to exploit the hybrid nature of the data acquired within the Pierre Auger Observatory. Handling all detector data within the same analysis software will make it easily possible to develop analysis procedures combining data from SD and radio, FD and radio or SD, FD and radio altogether. This will be possible both for measured data and for simulated events. Developing the ``radio-hybrid'' analysis strategies is yet a challenge for the future. The technical prerequisites for this development have, however, been successfully provided with the inclusion of radio functionality in the Offline framework.


\section{Example for an analysis pipeline} \label{sec:analysis}

To illustrate the flexibility and level of sophistication achieved with the radio functionality in Offline, we discuss here a module sequence processing \emph{simulated} data with a reconstruction pipeline that incorporates all relevant detector effects. Each line in the XML file listed in Fig.\ \ref{fig:modulesequence} denotes a module being invoked to perform a specific analysis step. Radio modules starting with \emph{RdChannel} manipulate the event data on the \emph{Channel} level, modules starting with \emph{RdStation} manipulate data on the \emph{Station} level. The modules starting with \emph{RdAntenna} perform the transitions between the two levels (cf.\ section \ref{sec:efieldreco}). Modules can easily be removed, replaced or rearranged to change the analysis procedure without having to recompile the source code. In the following, we will briefly discuss the individual steps of the module sequence and show how the signal evolves on its way through the analysis pipeline.\\

\begin{figure}[h!]
\begin{lstlisting}[style=xml_sty]
<module> EventFileReaderOG                    </module>
<module> RdStationAssociator                  </module>
<!--  cf. Fig. 6 raw simulated electric fields      -->

<module> RdAntennaStationToChannelConverter   </module>
<module> RdChannelNoiseGenerator              </module>
<!--  cf. Fig. 7 voltages at antenna foot-points    -->

<module> RdChannelResponseIncorporator        </module>
<!--  cf. Fig. 8 voltages at ADC                    -->

<module> RdChannelResampler                   </module>
<module> RdChannelTimeSeriesClipper           </module>
<module> RdChannelVoltageToADCConverter       </module>
<!--  signal is now equivalent to raw ADC values    -->

<module> RdChannelADCToVoltageConverter       </module>
<module> RdChannelPedestalRemover             </module>
<module> RdChannelResponseIncorporator        </module>
<module> RdChannelRFISuppressor               </module>
<module> RdChannelUpsampler                   </module>
<module> RdChannelBandpassFilter              </module>
<!--  cf. Fig. 9 voltages after signal processing   -->

<loop numTimes="unbounded">
  <module> RdAntennaChannelToStationConverter </module>
  <module> RdStationSignalReconstructor       </module>
  <module> RdDirectionConvergenceChecker      </module>
  <module> RdPlaneFit                         </module>
</loop>

<module> RdStationWindowSetter                </module>
<!--  cf. Fig. 10 field strengths in clipped trace  -->

<module> RdStationTimeSeriesWindower          </module>
<!--  cf. Fig. 11 Hann-windowed field strengths     -->

<module> RecDataWriter                        </module>
\end{lstlisting}
\caption{Example module sequence for performing a full detector simulation and event reconstruction of a simulated radio event. The time series data and frequency spectra extracted at certain steps during this analysis pipeline are illustrated with the figures referenced in the sequence. Analysis modules can be re-ordered or exchanged without any recompilation of the source code.}\label{fig:modulesequence}
\end{figure}

\subsection{Read-in and association}

\begin{figure*}[p]
\centering
\includegraphics[height=0.48\textwidth,angle=270]{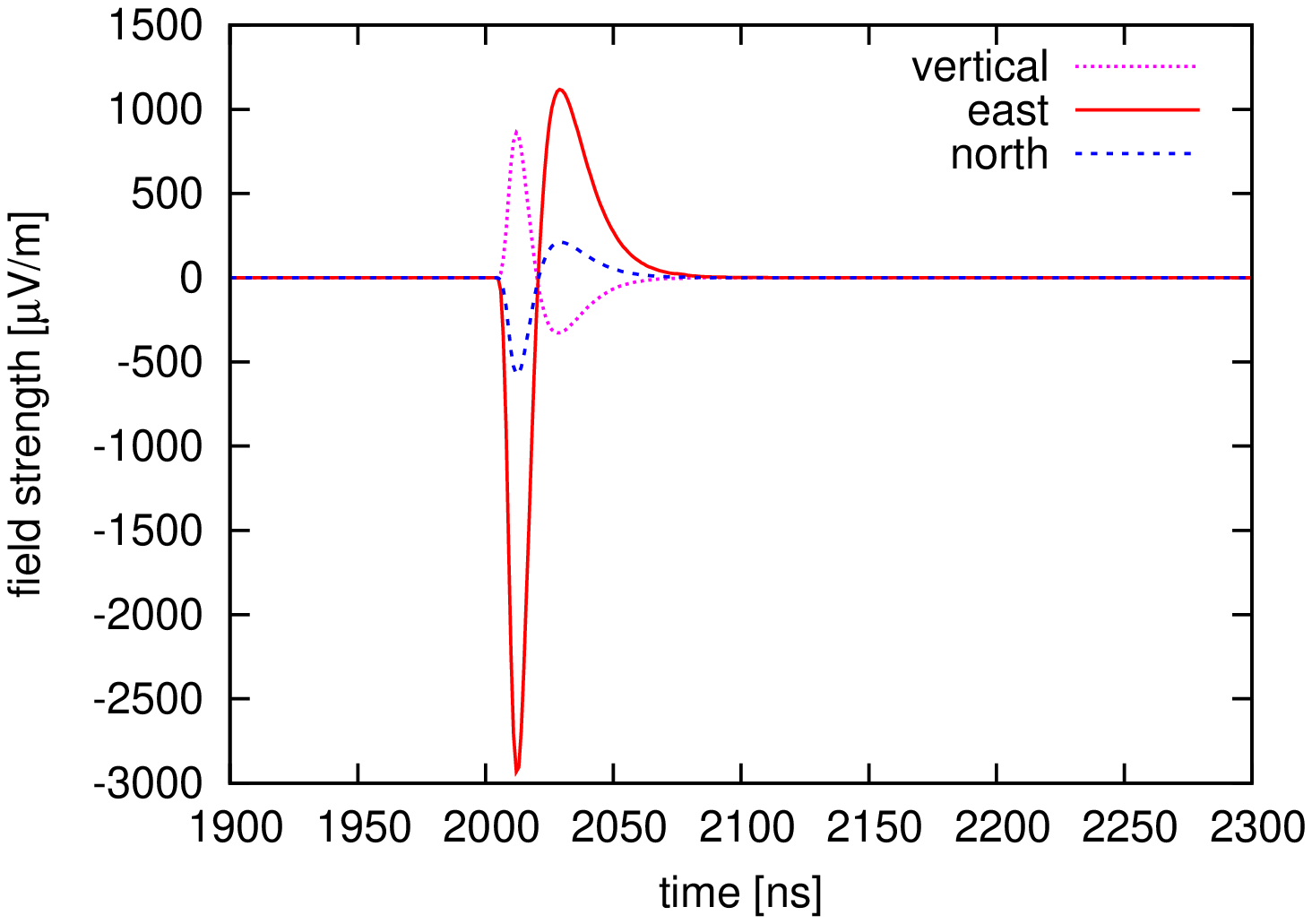}
\includegraphics[height=0.48\textwidth,angle=270]{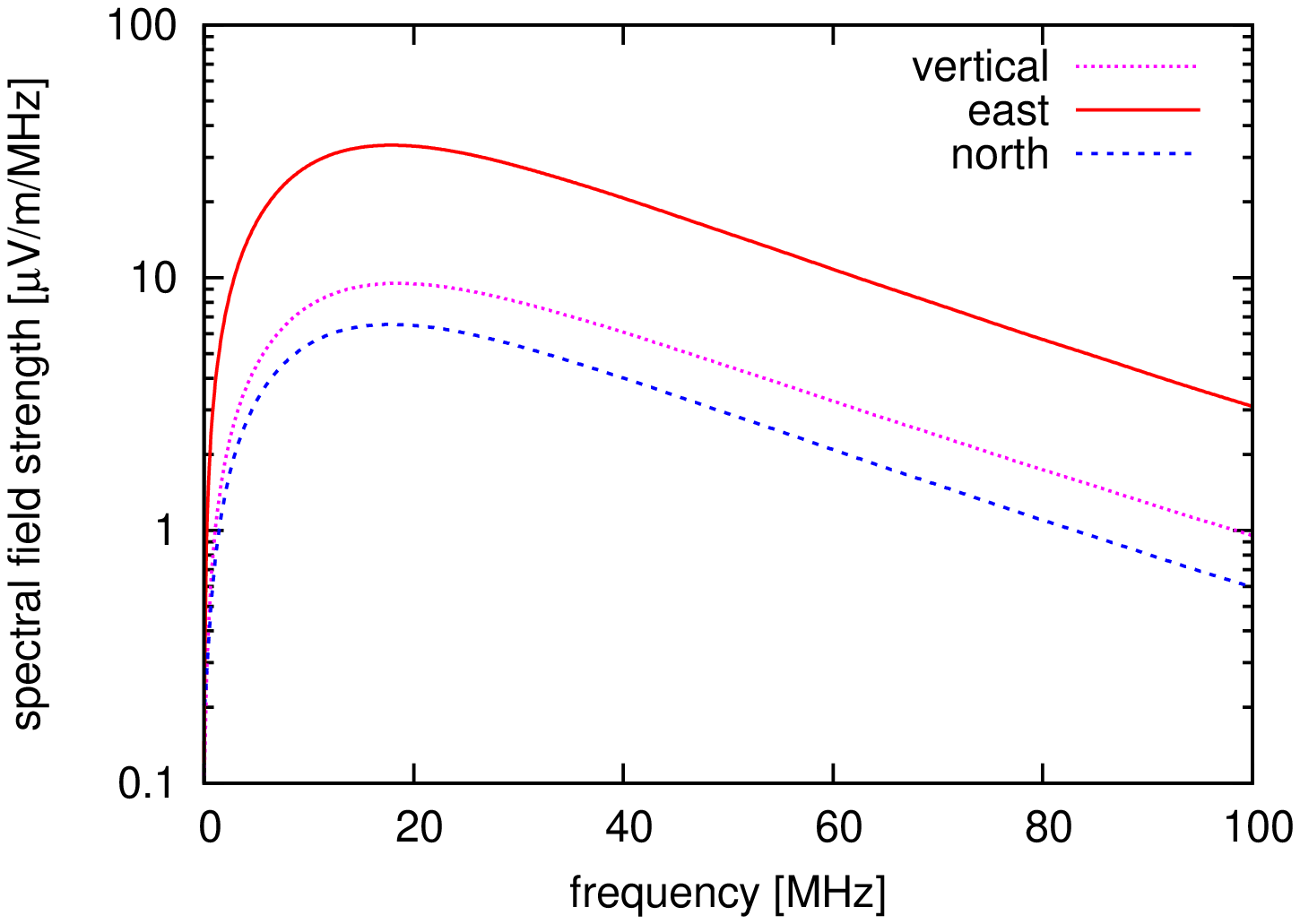}
\caption{Time traces (left) and frequency spectra (right) of a simulated event for the raw simulated three-dimensional electric field vector.} \label{fig:rawsim}
\end{figure*}

The module sequence starts with the read-in of simulated event data\footnote{The example event used here has an energy of $2.1\cdot10^{18}$~eV, a zenith angle of $58.4^{\circ}$, an azimuth angle of $291.0^{\circ}$ corresponding to an arrival direction of approximately SSW, and has been simulated with a proton as primary particle.} using the \texttt{EventFileReaderOG} module. After read-in, the simulated data represent an ``abstract'' simulation that is not yet associated to any detector stations. This association is performed by the \texttt{RdStationAssociatior} module, which associates the simulated signal traces with the corresponding stations in the field, and at the same time pads them appropriately to ensure that the signal falls into the correct part of the time series trace. Afterwards, the \emph{Station} data structure contains the physical electric field vector as predicted by the simulation, without the inclusion of any detector effects. The corresponding traces and spectra are depicted in Fig.\ \ref{fig:rawsim}.

\subsection{Simulation of the detector response}
 
\begin{figure*}[p]
\centering
\includegraphics[height=0.48\textwidth,angle=270]{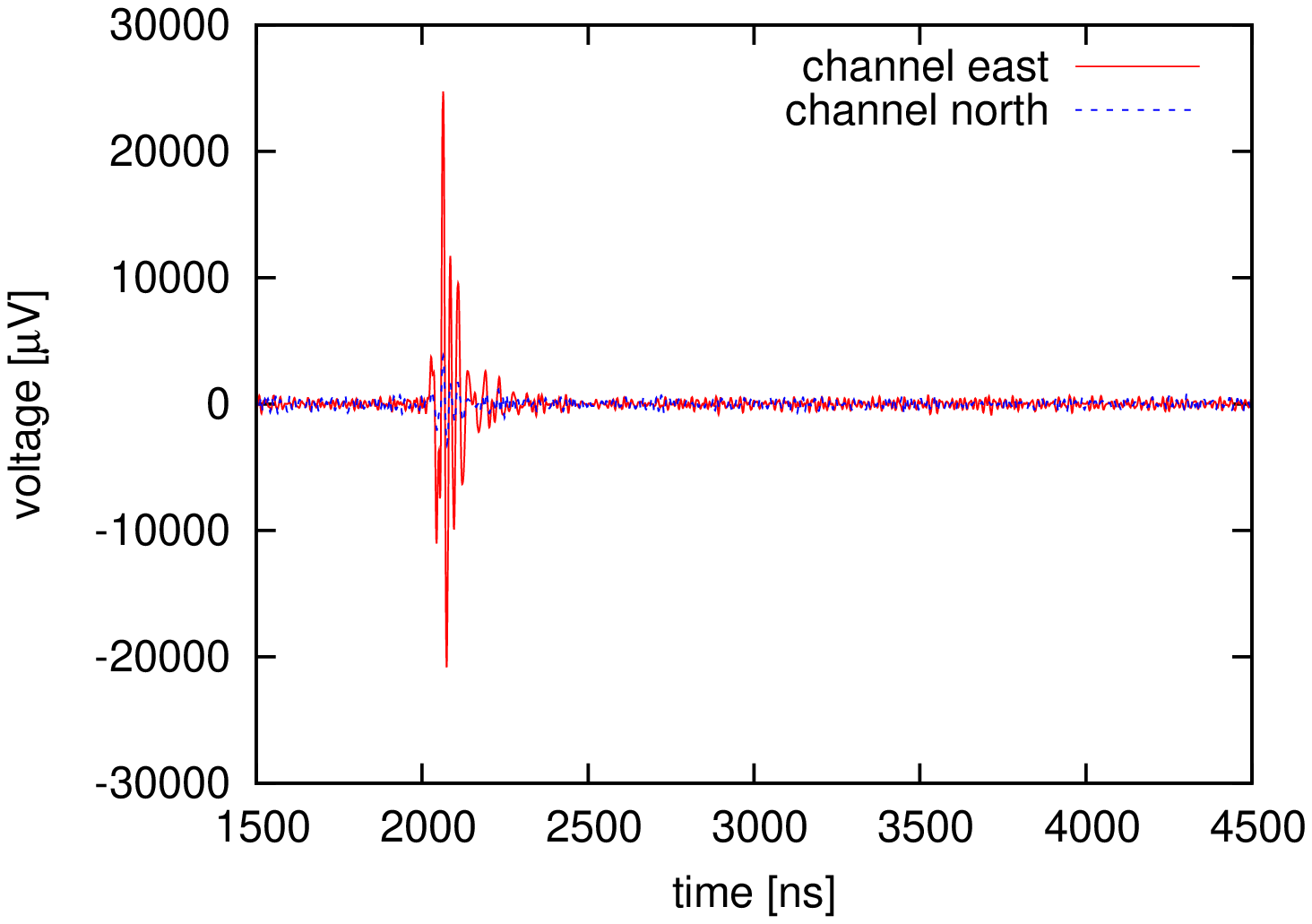}
\includegraphics[height=0.48\textwidth,angle=270]{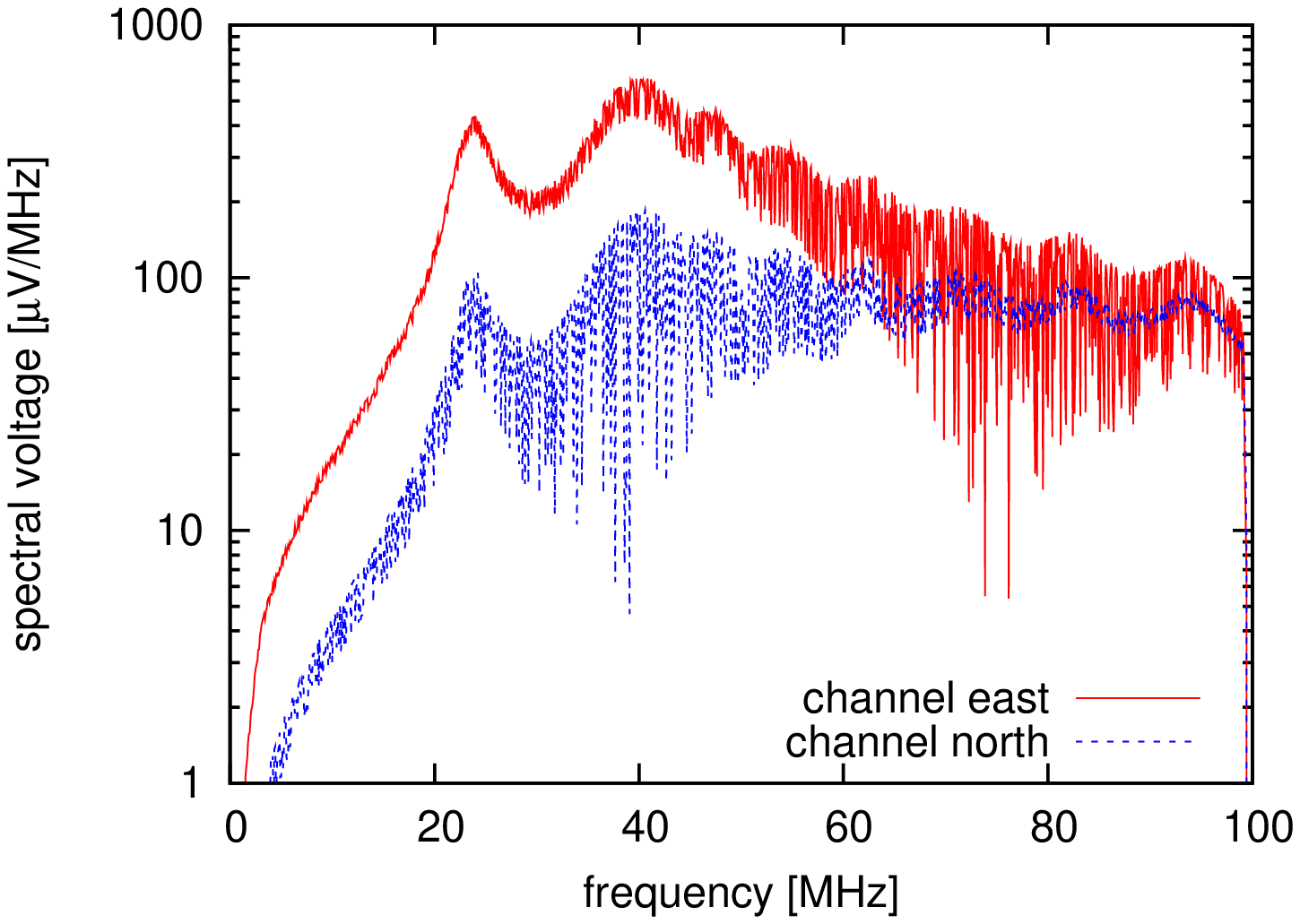}
\caption{Time traces (left) and frequency spectra (right) of the simulated event for the east and north channels. Using the simulated antenna characteristics (including the amplification by the LNA), the three-dimensional electric field vector has been projected to the two measurement channels. After the projection, white noise has been added by the \texttt{RdChannelNoiseGenerator}. The signal depicted here is what would be measured at the antenna foot-points over the whole frequency bandwidth. Please note that the frequency spectra correspond to the complete trace and not the zoomed-in time window shown here.} \label{fig:afterantenna}
\end{figure*}

The next steps in the module sequence change the data such that they become equivalent to data measured experimentally. The \texttt{RdAntennaStationToChannelConverter} calculates the signal voltages that each \emph{Channel} of a given \emph{Station} would have seen at the foot-points of the corresponding antennas by folding in the antenna response applicable to each individual channel. In the typical case of two antenna \emph{Channels} per \emph{Station}, this means that the three-dimensional electric field vector is projected to a two-dimensional surface. The \texttt{RdChannelNoiseGenerator} module then adds broad-band radio noise to the event. The resulting, simulated data for the east and north channels are shown in Fig.\ \ref{fig:afterantenna}.

\begin{figure*}[p]
\centering
\includegraphics[height=0.48\textwidth,angle=270]{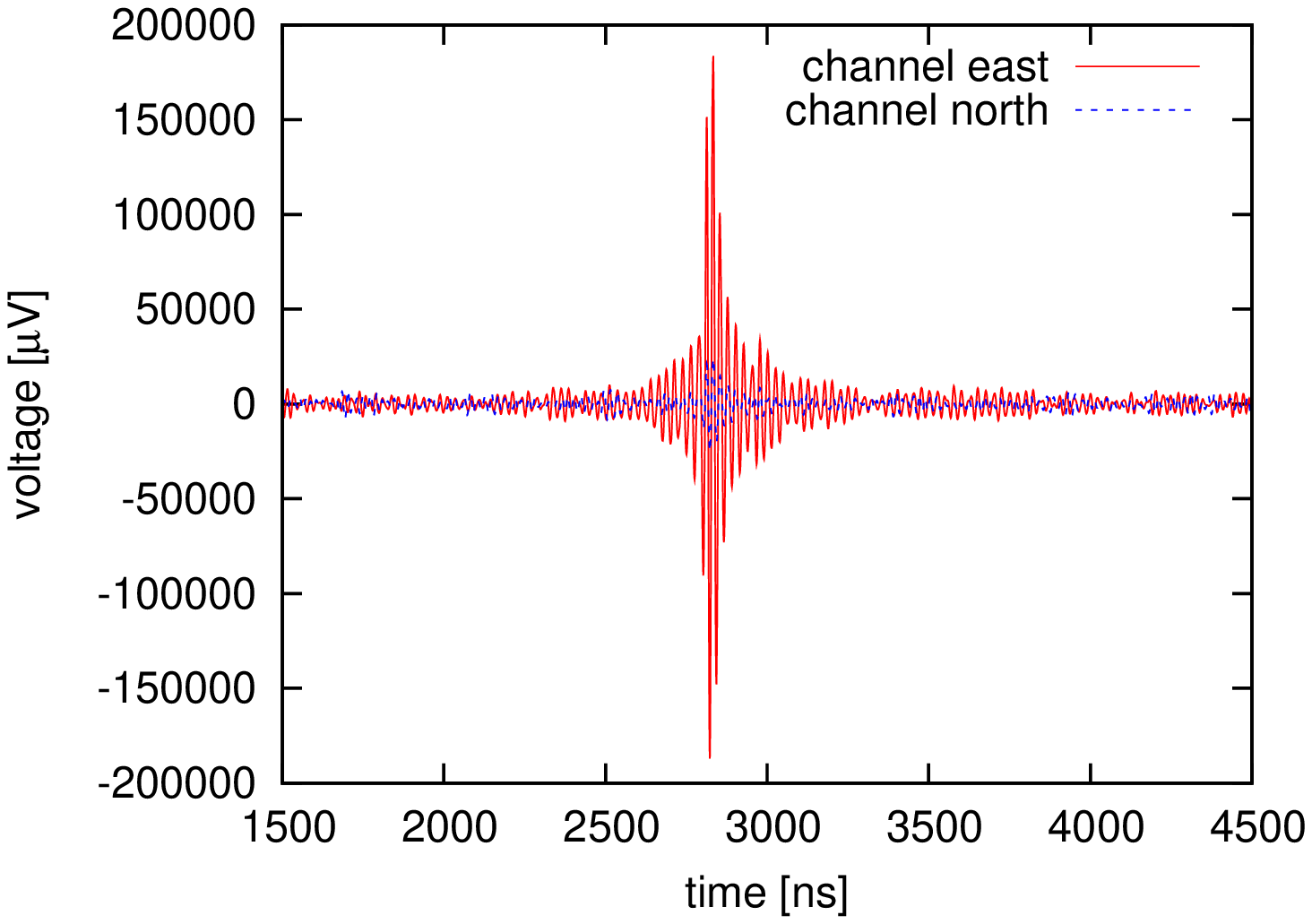}
\includegraphics[height=0.48\textwidth,angle=270]{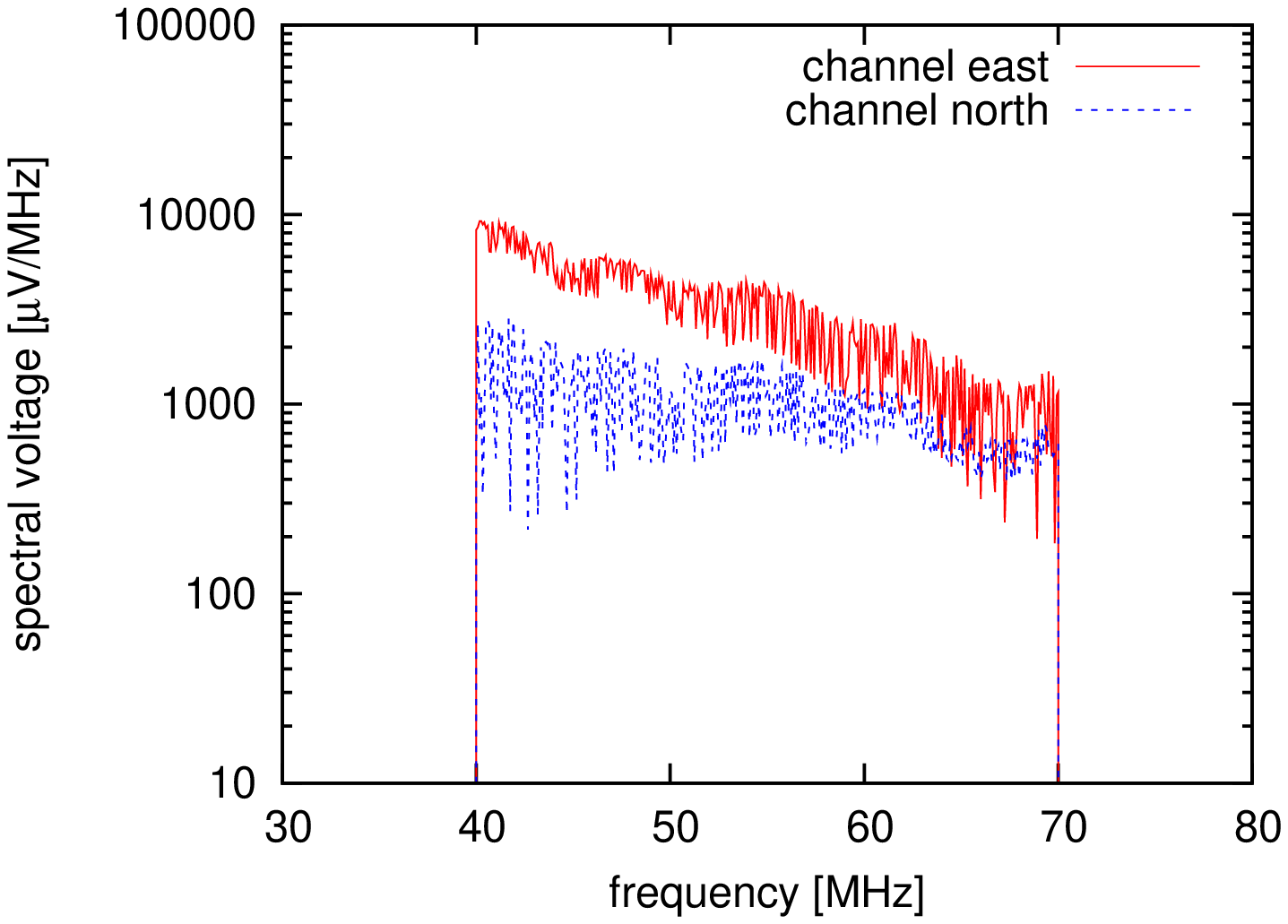}
\caption{Time traces (left) and frequency spectra (right) of the simulated event for the east and north channels after folding in the channel response (amplifiers, filters, cable). Note that the cable delays have shifted the time pulses to later times. Also, the spectral bandwidth has been limited to the design bandwidth of the experimental channels, leading to a broadening of the pulses. The signal depicted here is what would be measured at the channel ADCs. Please note that the frequency spectra correspond to the complete trace and not the zoomed-in time window shown here.} \label{fig:withchannelresponse}
\end{figure*}

The following call of the \texttt{RdChannelResponseIncorporator} incorporates the (forward) detector response of the cables, filters and amplifiers comprising the corresponding \emph{Channel}. After this module, the signal represents the voltages that would be measured at the channel ADCs, depicted in Fig.\ \ref{fig:withchannelresponse}.

The following steps convert this voltage at the ADCs to the signal that the channel ADC would indeed have measured. The \texttt{RdChannelResampler} module re-samples the \emph{Channel} time series data to the time-base with which the data are sampled in the experiment. (The prerequisite to this module is that high-frequency components which could lead to aliasing effects have been suppressed. This is ensured here because the \texttt{RdChannelResponseIncorporator} includes filters that do just that in the experimental setup.) The \texttt{RdChannelTimeSeriesClipper} then clips the \emph{Channel} traces to the number of samples which are taken in the experimental setup. Finally, the \texttt{RdChannelVoltageToADCConverter} converts the voltages of each sample to ADC counts that would have been recorded by the channel ADCs, thereby taking into account quantization and saturation effects.

At this point, all of the relevant detector effects have been incorporated in the simulated traces. In other words, the data now have the same properties as measured data directly after read-in. Consequently, the remainder of the reconstruction pipeline is identical to one that would be applied directly to measured data. Choosing the reconstruction procedure applied to the simulated data identical to the one applied to measured data makes sure that even subtle changes introduced by individual analysis modules can be investigated on the basis of simulations.

\subsection{Signal cleaning}

After converting the ADC counts back to voltages with the \texttt{RdChannelADCToVoltageConverter} and removing a possible DC offset of the ADC with the \texttt{RdChannelPedestalRemover}, the characteristics of the analogue components of each \emph{Channel} are folded out from the data with a second call of the \texttt{RdChannelResponseIncorporator}.

\begin{figure*}[p]
\centering
\includegraphics[height=0.48\textwidth,angle=270]{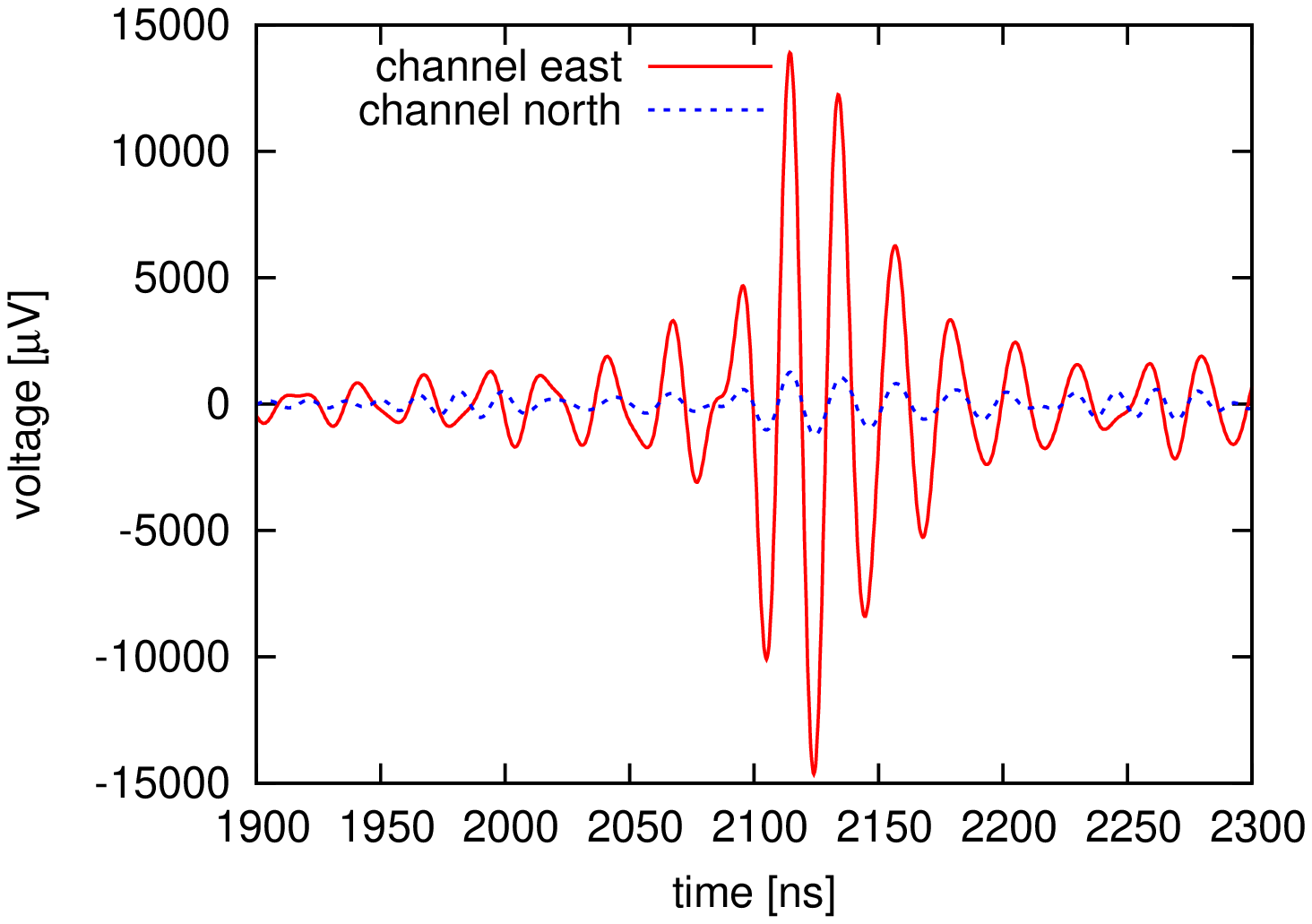}
\includegraphics[height=0.48\textwidth,angle=270]{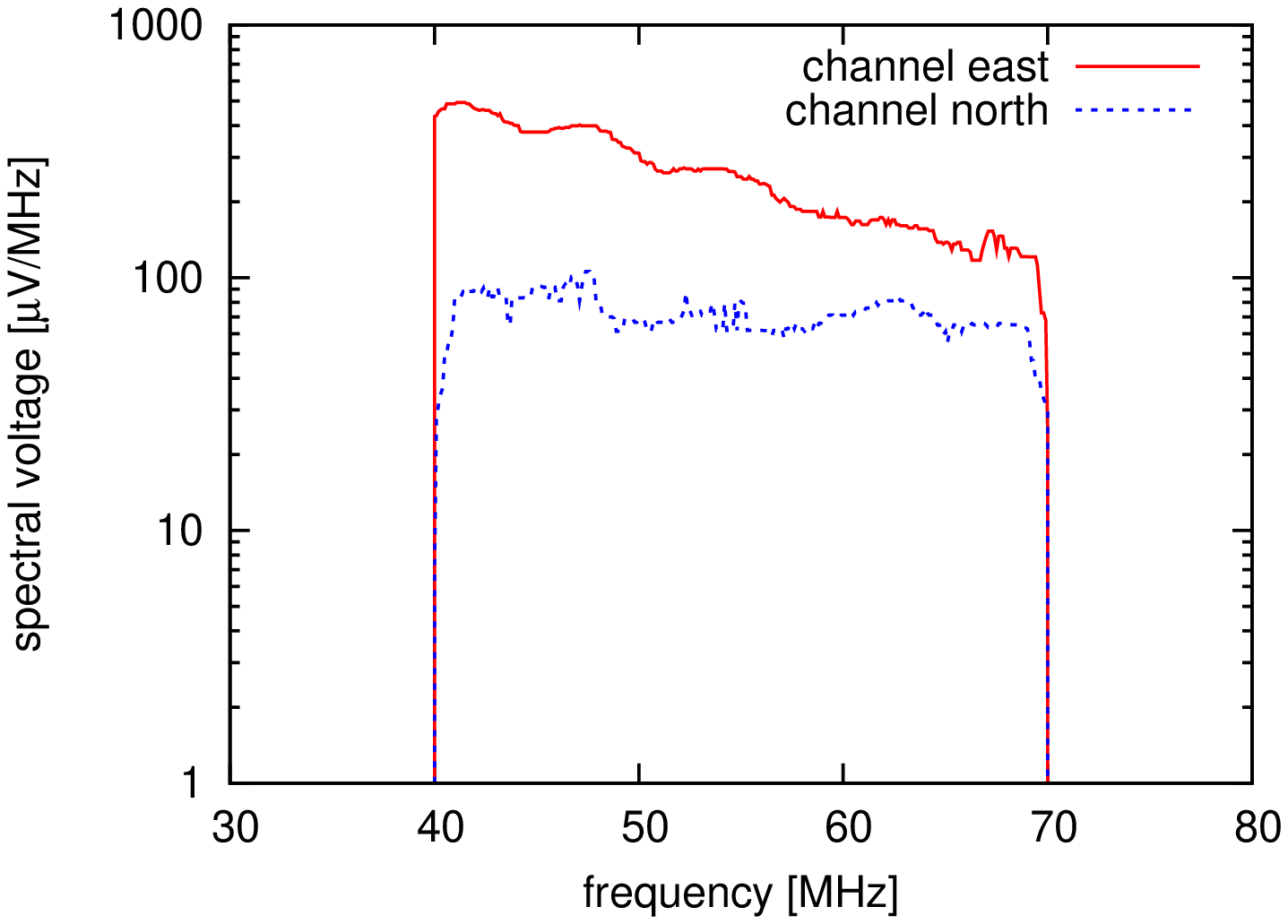}
\caption{Time traces (left) and frequency spectra (right) of the simulated event after all processing steps on the \emph{Channel} level. The smoothing in the frequency spectra is due to the \texttt{RdChannelRFISuppressor}. These data again correspond to the signal measured at the foot-points of the antennas, but this time limited to the 40-70~MHz band. They are the starting point for the reconstruction of the three-dimensional electric field vector. Please note that the frequency spectra correspond to the complete trace and not the zoomed-in time window shown here.} \label{fig:simbeforereconstruction}
\end{figure*}

The following steps are intended to improve the reconstruction quality by the use of advanced digital processing techniques. In a first step, the \texttt{RdChannelRFISuppressor} module suppresses narrow-band signals (e.g., TV carriers present in measured data) using a median filter. It is followed by the \texttt{RdChannelUpsampler} module which correctly reconstructs (interpolates) the signal on a finer time-base. (This is possible if the complete signal information is present in the digitized signal, i.e., if the Nyquist criterion for data sampling is fulfilled.) In a further step the signal bandwidth is limited by a digital bandpass filter using the \texttt{RdChannelBandpassFilter} module. After these steps, the signal is ready for the reconstruction of the three-dimensional electric field vector and looks like the data presented in Fig.\ \ref{fig:simbeforereconstruction}.

\subsection{Vectorial reconstruction}

The following loop performs an iterative reconstruction of the three-dimensional electric field vector and the signal arrival direction. The \texttt{RdAntennaChannelToStationConverter} performs the reconstruction of the three-dimensional electric field described in section \ref{sec:efieldreco}. Afterwards, the \texttt{RdStationSignalReconstructor} identifies the times at which radio pulses have been detected and quantifies the parameters of these pulses. Next, the \texttt{RdPlaneFit} reconstructs the arrival direction of the radio signal with a plane-wave assumption based on the previously established pulse arrival times. Finally, the \texttt{RdDirectionConvergenceChecker} module tests whether the iterative procedure has converged or not and breaks the loop accordingly.

\subsection{Post-processing}

\begin{figure*}[p]
\centering
\includegraphics[height=0.48\textwidth,angle=270]{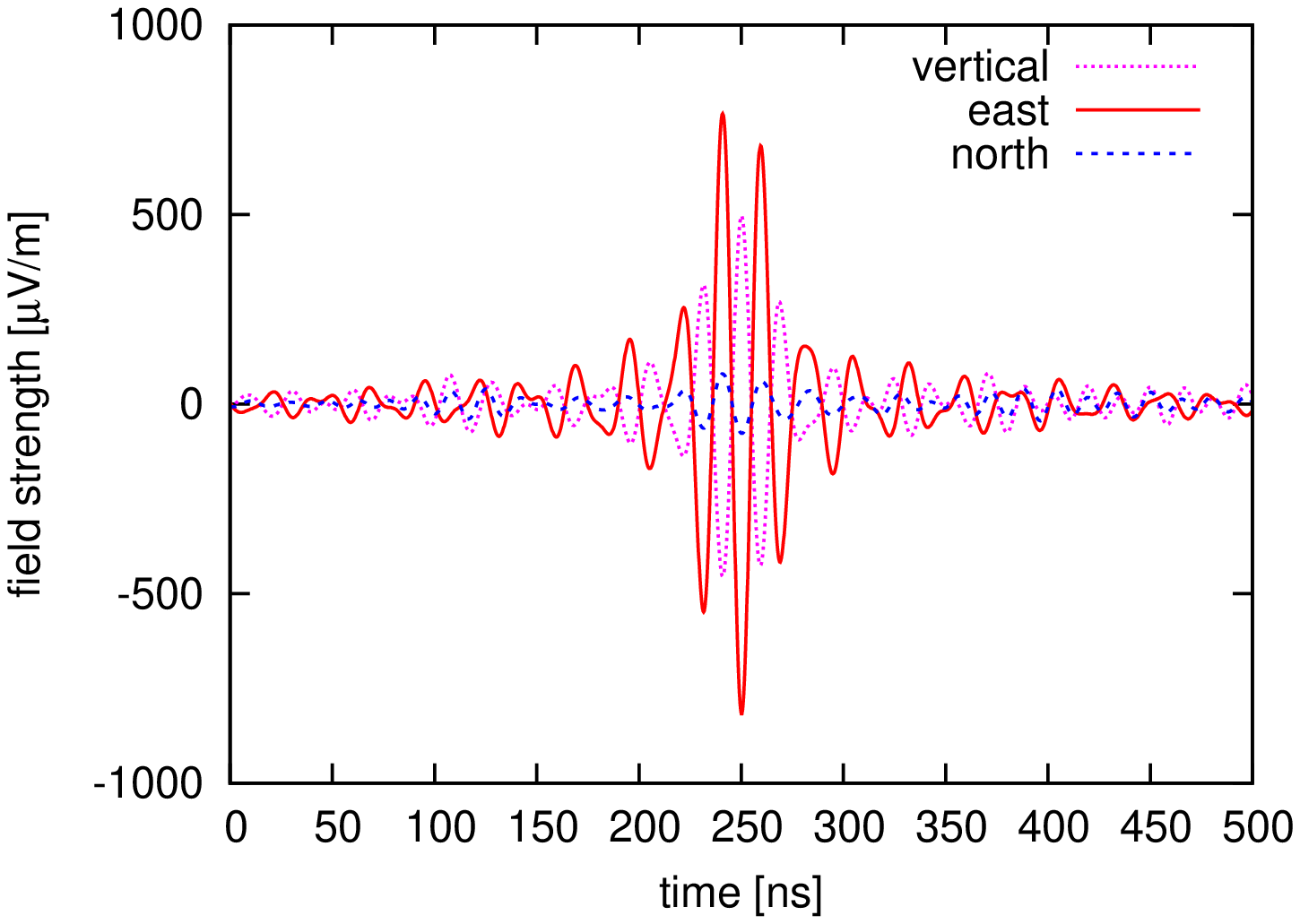}
\includegraphics[height=0.48\textwidth,angle=270]{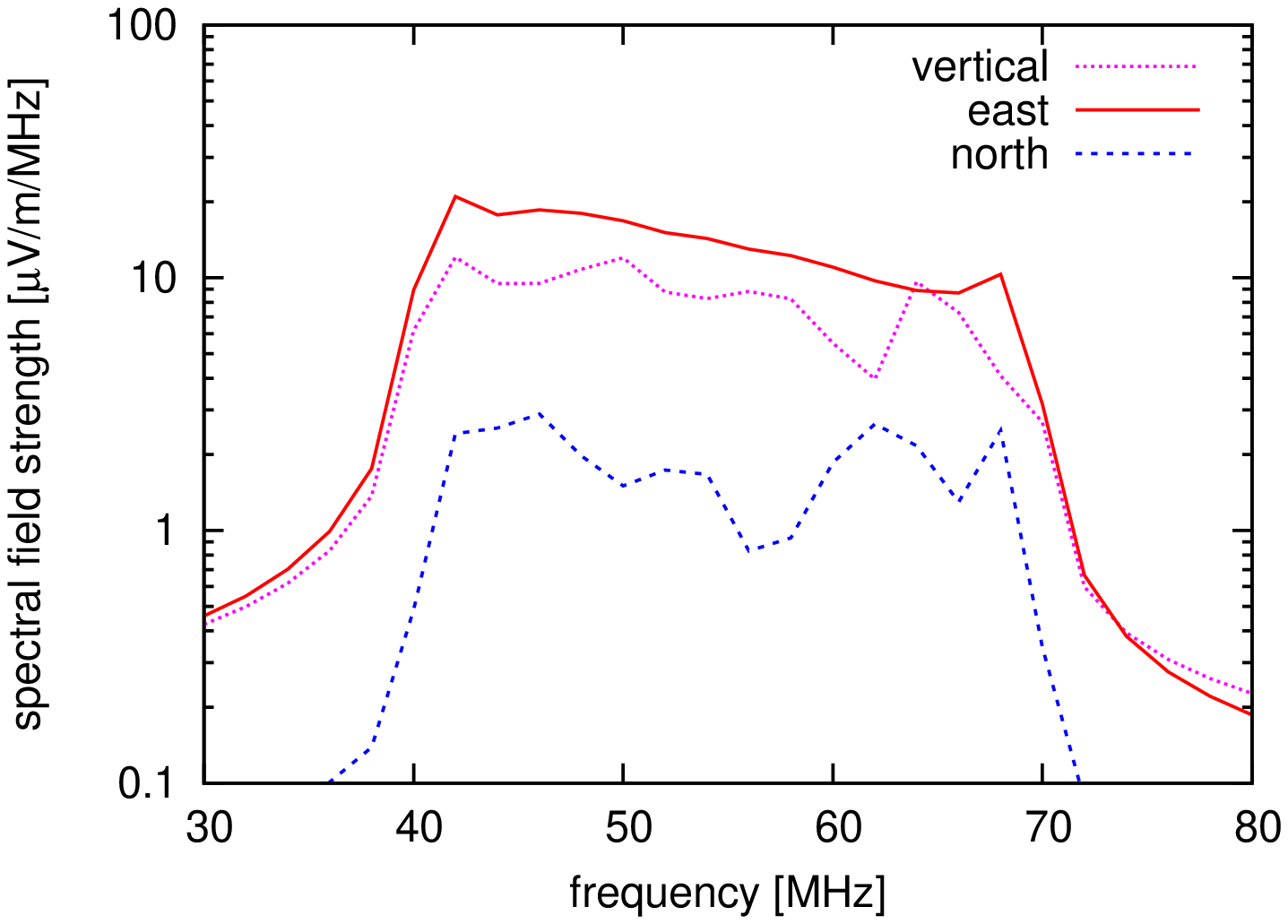}
\caption{Time traces (left) and frequency spectra (right) for the reconstructed three-dimensional electric field vector of the simulated event after application of the \texttt{RdChannelWindowSetter}. The leakage visible in the frequency spectra is due to the cutting of the time traces to a 500~ns window around the detected pulses. The frequency spectra correspond to the time traces shown here.} \label{fig:simleakage}
\end{figure*}

\begin{figure*}[p]
\centering
\includegraphics[height=0.48\textwidth,angle=270]{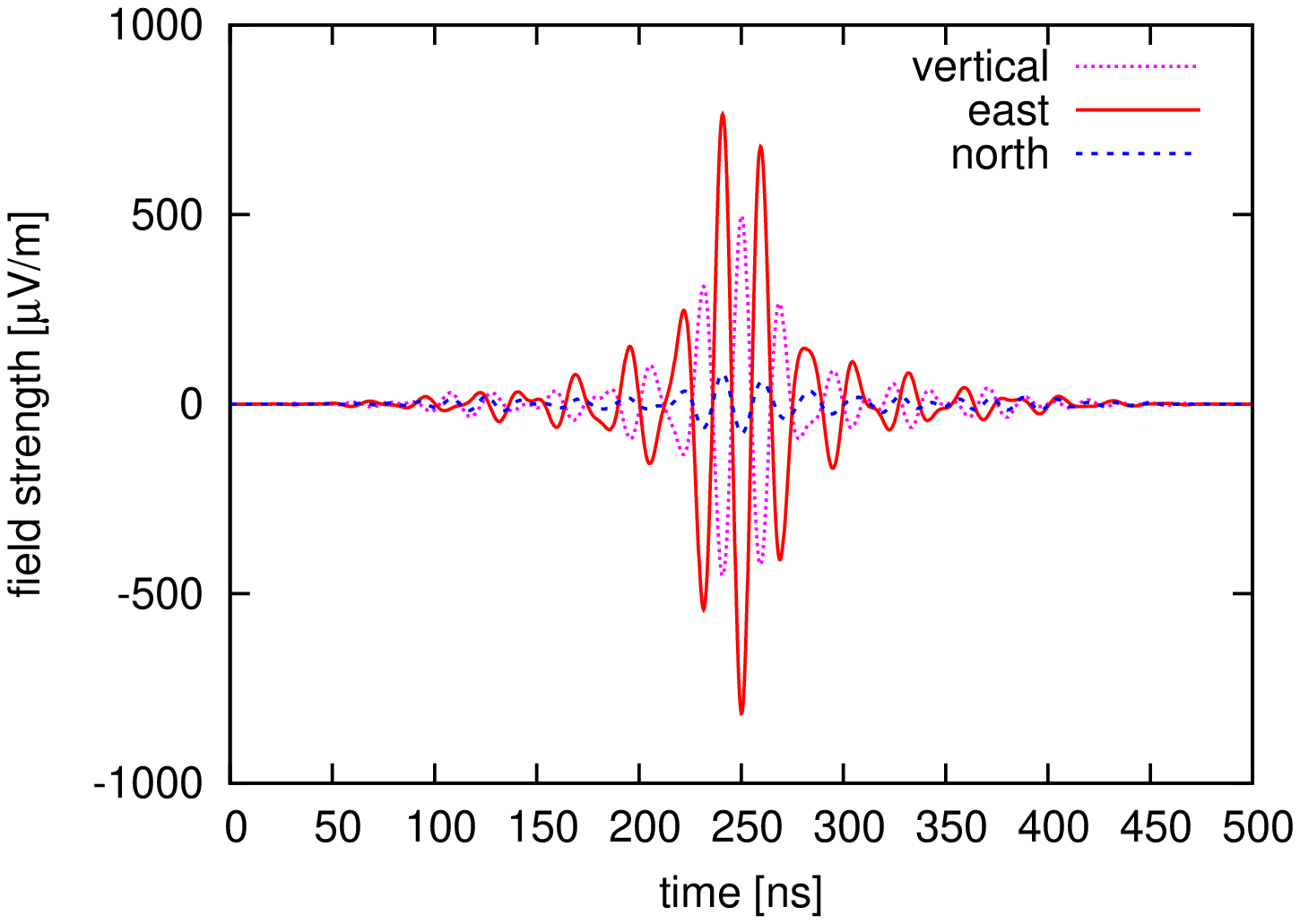}
\includegraphics[height=0.48\textwidth,angle=270]{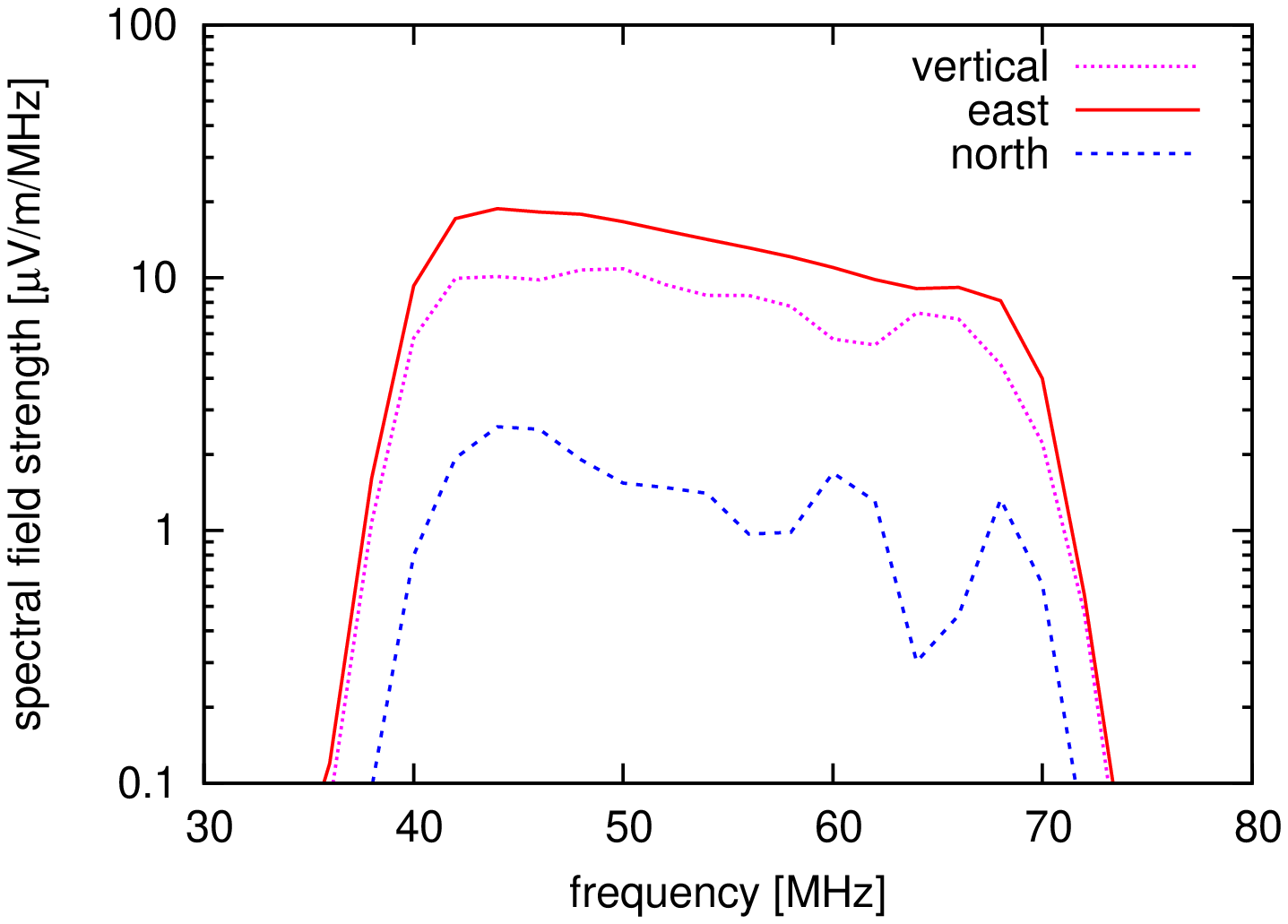}
\caption{Time traces (left) and frequency spectra (right) of the simulated event for the final, Hann-windowed three-dimensional electric field vector. The significant vertical component of the raw simulation has been reconstructed by the analysis chain. The final reconstructed arrival direction is $\theta = 60.7^{\circ}$ and $\phi = 295.1^{\circ}$. The input direction for the simulated event was $\theta = 58.4^{\circ}$ and $\phi = 291.0^{\circ}$. The frequency spectra correspond to the time traces shown here.} \label{fig:simfinal}
\end{figure*}

After breaking the iterative reconstruction loop, the vectorial, detector-independent electric field has been completely reconstructed. For practical purposes, the time series is then restricted to a window of 500~ns around the detected pulses with the \texttt{RdStationWindowSetter}. This leads to leakage effects in the frequency spectra, visible in Fig.\ \ref{fig:simleakage}. To suppress these leakage artifacts, a Hann window is applied with the \texttt{RdStationTimeSeriesWindower}. The final reconstructed signal is then seen in Fig.\ \ref{fig:simfinal}. As a last step, the \texttt{RecDataWriter} call writes out the reconstructed event data to an ADST file for further processing in higher-level analyses.

\newpage

\newpage

\section{Conclusions and Outlook}

We have implemented a complete set of radio analysis functionality in the Offline software framework of the Pierre Auger Observatory. The radio functionality has been included in a canonical and seamless way in addition to the existing SD and FD functionality. This approach will make the realization of ``radio-hybrid'' analysis strategies in the future straight-forward.

Already now, however, the radio functionality in Offline has reached a high degree of sophistication with highlights such as a very fine-grained simulation of detector effects, advanced signal processing algorithms, transparent and efficient handling of FFTs, read-in of multiple file formats for measured and simulated radio data, and in particular the reconstruction of the three-dimensional electric field vector from two-dimensional measurements. Planned improvements encompass the implementation of a curved fit, inclusion of interferometric radio analysis functionality, and the handling of a time-variable detector including a fine-grained treatment of the instrumental calibration.

Parties interested in using the functionality are encouraged to contact the corresponding author. The source code can be made available on request.

\section*{Acknowledgements}

We would like to thank our colleagues from the LOPES collaboration for providing the source code of their analysis software openly to the public \citep{LofarSoft}. Many algorithms for radio analysis and reconstruction have been inspired by or based on those used within LOPES. This research has been supported by grant number VH-NG-413 of the Helmholtz Association.

The successful installation and commissioning of the Pierre Auger Observatory
would not have been possible without the strong commitment and effort
from the technical and administrative staff in Malarg\"ue.

We are very grateful to the following agencies and organizations for financial support: 
Comisi\'on Nacional de Energ\'{\i}a At\'omica, 
Fundaci\'on Antorchas,
Gobierno De La Provincia de Mendoza, 
Municipalidad de Malarg\"ue,
NDM Holdings and Valle Las Le\~nas, in gratitude for their continuing
cooperation over land access, Argentina; 
the Australian Research Council;
Conselho Nacional de Desenvolvimento Cient\'{\i}fico e Tecnol\'ogico (CNPq),
Financiadora de Estudos e Projetos (FINEP),
Funda\c{c}\~ao de Amparo \`a Pesquisa do Estado de Rio de Janeiro (FAPERJ),
Funda\c{c}\~ao de Amparo \`a Pesquisa do Estado de S\~ao Paulo (FAPESP),
Minist\'erio de Ci\^{e}ncia e Tecnologia (MCT), Brazil;
AVCR, AV0Z10100502 and AV0Z10100522,
GAAV KJB300100801 and KJB100100904,
MSMT-CR LA08016, LC527, 1M06002, and MSM0021620859, Czech Republic;
Centre de Calcul IN2P3/CNRS, 
Centre National de la Recherche Scientifique (CNRS),
Conseil R\'egional Ile-de-France,
D\'epartement  Physique Nucl\'eaire et Corpusculaire (PNC-IN2P3/CNRS),
D\'epartement Sciences de l'Univers (SDU-INSU/CNRS), France;
Bundesministerium f\"ur Bildung und Forschung (BMBF),
Deutsche Forschungsgemeinschaft (DFG),
Finanzministerium Baden-W\"urttemberg,
Helmholtz-Gemeinschaft Deutscher Forschungszentren (HGF),
Ministerium f\"ur Wissenschaft und Forschung, Nordrhein-Westfalen,
Ministerium f\"ur Wissenschaft, Forschung und Kunst, Baden-W\"urttemberg, Germany; 
Istituto Nazionale di Fisica Nucleare (INFN),
Istituto Nazionale di Astrofisica (INAF),
Ministero dell'Istruzione, dell'Universit\`a e della Ricerca (MIUR), 
Gran Sasso Center for Astroparticle Physics (CFA), Italy;
Consejo Nacional de Ciencia y Tecnolog\'{\i}a (CONACYT), Mexico;
Ministerie van Onderwijs, Cultuur en Wetenschap,
Nederlandse Organisatie voor Wetenschappelijk Onderzoek (NWO),
Stichting voor Fundamenteel Onderzoek der Materie (FOM), Netherlands;
Ministry of Science and Higher Education,
Grant Nos. 1 P03 D 014 30 and N N202 207238, Poland;
Funda\c{c}\~ao para a Ci\^{e}ncia e a Tecnologia, Portugal;
Ministry for Higher Education, Science, and Technology,
Slovenian Research Agency, Slovenia;
Comunidad de Madrid, 
Consejer\'{\i}a de Educaci\'on de la Comunidad de Castilla La Mancha, 
FEDER funds, 
Ministerio de Ciencia e Innovaci\'on and Consolider-Ingenio 2010 (CPAN),
Generalitat Valenciana, 
Junta de Andaluc\'{\i}a, 
Xunta de Galicia, Spain;
Science and Technology Facilities Council, United Kingdom;
Department of Energy, Contract Nos. DE-AC02-07CH11359, DE-FR02-04ER41300,
National Science Foundation, Grant No. 0969400,
The Grainger Foundation USA; 
ALFA-EC / HELEN,
European Union 6th Framework Program,
Grant No. MEIF-CT-2005-025057, 
European Union 7th Framework Program, Grant No. PIEF-GA-2008-220240,
and UNESCO.








\end{document}